\documentclass[final]{ws-p8-50x6-00}
\usepackage{amssymb,amsmath,epsfig}

\begin{document}
\title{Computational probes of  Collective excitations in low-dimensional magnetism}
\author{Gerhard M{\"u}ller} 
\address{Department of Physics, University of Rhode Island,
         Kingston RI 02881, USA\\
         Email: gmuller@uri.edu}
\author{Michael Karbach} 
\address{Bergische Universit{\"a}t Wuppertal, Fachbereich Physik, 
  42097 Wuppertal, Germany\\
  Email: michael@karbach.org}

\maketitle
\abstracts{The investigation of the dynamics of quantum many-body systems is a
  concerted effort involving computational studies of mathematical models and
  experimental studies of material samples. Some commonalities of the two tracks
  of investigation are discussed in the context of the quantum spin dynamics of
  low-dimensional magnetic systems, in particular spin chains. The study of
  quantum fluctuations in such systems at equilibrium amounts to exploring the
  spectrum of collective excitations and the rate at which they are excited
  from the ground state by dynamical variables of interest.  The exact results
  obtained  via Bethe ansatz or  algebraic analysis (quantum groups) for a
  select class of completely integrable models can be used as benchmarks
  for numerical studies of nonintegrable models, for which computational access
  to the spectrum of collective excitations is limited.}
%
\section{Introduction}\label{sec:1}
%
Notwithstanding the fact that experimental and theoretical studies of condensed
matter systems are fundamentally complementary to each other, they share
important features, which we wish to illuminate here in the context of quantum
many-body dynamics.  The object of study is a {\it sample} for {\it experimental
  probes} and a {\it model} for {\it computational probes}.  Whether the model
is regarded as a {\it mathematical idealization} of a real chunk of matter or
whether the sample is viewed as a {\it physical realization} of a system defined
with mathematical rigor may be a matter of perspective, but the influence
exerted by theory and experiment is mutual. Any \textit{observation} of note
calls for an \textit{explanation}, while any \textit{prediction} of substance
brings into motion attempts at \textit{verification}.

When performing experimental or computational probes in quantum many-body
dynamics, the goal of researchers is very much the same, namely trying to make
sense of what, in general, is a jumble of fluctuations to which the fundamental
degrees of freedom contribute in various combinations and configurations.
Experimental probes begin with a \textit{measurement} and computational probes
with a \textit{calculation}. The results of either probe then require an
\textit{interpretation} in terms and concepts that are common to both approaches.

Qualitatively, the specification of a quantum many-body system involves
information on {\it composition}, {\it interaction}, and {\it environment}.
Theoretically this information is encoded in the {\it Hamiltonian} and in the
{\it density operator}, whereas experimentally part of this information is
manifest or hidden in the {\it sample} and other parts are controlled by the {\it
  instrumental setup}.

Any system thus specified is subject to fluctuations, of which we distinguish
three kinds. Depending on the circumstances, each kind may govern the dynamical
properties of the system. {\it Thermal fluctuations} are likely to be dominant
in any statistical ensemble at elevated temperatures. {\it Parametric
  fluctuations} are manifestations of quenched random inhomogeneities in
composition (e.g. dilution) or interaction (e.g. random bonds or fields),
leaving distinct marks in dynamical properties, especially at low temperatures.
 
In the absence of thermal and parametric fluctuations, {\it quantum
  fluctuations} remain present. They are a direct consequence of the
(autonomous) quantum time evolution and the time-delayed projections that are
part of any dynamical probe, be it experimental or computational. No zero point
motion exists in classical Hamiltonian systems. Here the ground state
corresponds to a stable fixed point in phase space, and all dynamical variables
are constant.

Consider a (non-random) quantum many-body system in a pure quantum state. The
quantum fluctuations then depend on three quantities: (i) the Hamiltonian $H$,
(ii) the state, which we take here to be the ground state $|G\rangle$, and (iii) some
dynamical variable, here expressed by operator $A$.  The Hamiltonian governs the
dynamics deterministically for as long as the system can be regarded in
isolation. It makes no difference whether this time evolution is viewed in the
Schr{\"o}dinger or Heisenberg representation:
\begin{eqnarray}
  \label{eq:1}
  i\hbar\frac{d}{dt}|\psi\rangle = 
H|\psi\rangle\quad & \Longrightarrow & \quad |\psi(t)\rangle = e^{-iHt/ \hbar}|\psi\rangle, \quad |\psi\rangle \equiv A|G\rangle  \\
 i\hbar\frac{dA}{dt} = [A,H]\quad &\Longrightarrow & \quad A(t) = e^{iHt/ \hbar}Ae^{-iHt/ \hbar}.
\end{eqnarray}
The statistical element is introduced when we subject the time evolved state
$|\psi(t)\rangle$ or, equivalently, the time evolved dynamical variable $A(t)$, to an
observation. In practice, this step involves the observation or evaluation of a
dynamic correlation function, which can then be viewed either as the projection
of the time evolved state $|\psi(t)\rangle$ or  as the product of the dynamical
variable $A(t)$ at two different times evaluated in the stationary state $|G\rangle$:
\begin{equation}
  \label{eq:2}
  \langle A(t)A\rangle \stackrel{\rm
  equil.}{=} \langle AA(-t)\rangle = 
  \begin{cases}
    e^{iE_Gt/ \hbar}\langle\psi(0)|\psi(t)\rangle \\
    \langle G|Ae^{-iHt/ \hbar}Ae^{iHt/ \hbar}|G\rangle.
  \end{cases}
\end{equation}

The sum total of quantum fluctuations in a typical many-body system contains
many different dynamical modes no matter in which state the system happens to
be.  Only once we pick a set of dynamical variables (one variable at a time) do
we begin to gain insight into the role of specific modes taken from a multitude
of quantum fluctuations. Criteria for choosing dynamical variables include
experimental accessibility, computational amenability, or any specific purpose
like the study of ordering tendencies or phase transitions.

On the microscopic level, dynamical probes are bound to be somewhat
heavy-handed. No observation without perturbation! However, in all situations
considered here, we shall assume that the interaction between the probe and the
sample takes place in the regime of {\it linear response}.\cite{BH70}
This assumption is quite realistic for neutron scattering under most
circumstances.  Consider the weakly perturbed Hamiltonian $H(t)=H_0-b(t)B$,
where the external field $b(t)$ couples to the dynamical variable $B$ of the
system $H_0$.  The linear response of any other dynamical variable $A$ to that
perturbation,
\begin{equation}\label{eq:linres}
\langle A(t)\rangle-\langle A\rangle_0 = 
\int_{-\infty}^{+\infty}dt'\tilde{\chi}_{AB}(t-t')b(t'),
\end{equation}
is then determined by Kubo's formula for the response function,
\begin{equation}\label{eq:kubfor}
\tilde{\chi}_{AB}(t-t') = \frac{i}{\hbar}\Theta(t-t')\langle[A(t),B(t')]\rangle_0.
\end{equation}
Its Fourier transform, the generalized susceptibility
$\chi_{AB}(\omega+i\epsilon)$, has (for $\epsilon\to0$) a symmetric real part
and an antisymmetric imaginary part: $\chi_{AB}(\omega) = \chi'_{AB}(\omega) +
i\chi''_{AB}(\omega)$. The latter is related to the Fourier transform of the
correlation function $\langle A(t)A\rangle$, named the structure function
\begin{equation}\label{eq:fludis}
S_{AA}(\omega) \equiv 
\int_{-\infty}^{+\infty}dt\,e^{i\omega t}\langle A(t)A\rangle = 
\frac{2\hbar\chi''_{AA}(\omega)}{1-e^{-\beta\hbar\omega}},
\end{equation}
via the fluctuation-dissipation theorem.  The spectral representation of this
function, in particular the expression resulting in the limit $T\to0$, where
thermal fluctuations are absent,
\begin{equation}\label{eq:specrep}
S_{AA}(\omega) \stackrel{T=0}{=} 
2\pi\sum_{\lambda}|\langle G|A|\lambda\rangle|^2\delta
\left(\omega - (E_\lambda-E_G)/\hbar\right),
\end{equation}
demonstrates  that observing quantum fluctuations in the linear
response regime means observing collective modes and their transition rates from
the ground state. In the following, we look at quantum fluctuations and the
associated collective modes from this particular angle for a number of
situations as investigated by a variety of methods.\footnote{From here on we set 
  $\hbar=1$ to simplify the notation.}
%
\section{Magnons Excited from  Valence-Bond-Solid State}\label{sec:2}
%
Consider the one-dimensional (1D) $s=1$ Heisenberg model with bilinear and
biquadratic exchange
\begin{equation}\label{eq:Hbibi}
       H_\theta = J\sum_{n=1}^N \left[ 
         \cos \theta ({\bf S}_n \cdot {\bf S}_{n+1}) + 
         \sin \theta \left( {\bf S}_n \cdot {\bf S}_{n+1} \right)^2\right].
\end{equation}
At $T\!=\!0$ it has a phase with ferromagnetic long-range order (LRO)
($-\pi\!<\!\theta\!<\! -3\pi/4$ or $\pi/2\!<\!\theta\!\leq\!\pi$), a phase with dimer LRO
($-3\pi/4\!<\!\theta\!<\! -\pi/4$), the Haldane phase with hidden topological LRO
($-\pi/4\!<\!\theta\!<\!\pi/4$), and an obscure phase ($\pi/4<\theta<\pi/2$) that was
named trimerized.\cite{SJG96} These  ordering tendencies
help us choose suitable dynamical variables when we explore the spectrum of
collective excitations. All
dynamical variables used for this purpose will have the form of {\it fluctuation
  operators},
\begin{equation}\label{eq:FqA}
        F_Q^A \equiv \frac{1}{\sqrt{N}}\sum_{n=1}^Ne^{iQn}A_n,
\end{equation}
where the operator $A_n$ acts locally at or near lattice site $n$.  The
associated structure function (\ref{eq:specrep}) is called the {\it dynamic
  structure factor}:
\begin{equation}\label{eq:SAAqw}
  S_{AA}(Q,\omega) \equiv 
  \int_{-\infty}^{+\infty} dt\,e^{i\omega t} 
  \langle F_Q^A(t)F_Q^{A^\dagger} \rangle\stackrel{T=0}{=} 
  \sum_\lambda W_\lambda^A \delta(\omega-\omega_\lambda),
\end{equation} 
where the sum runs over the dynamically relevant excitations $|\lambda\rangle$
with energy $\omega_\lambda=E_\lambda-E_G$ and spectral weight
$W_\lambda^A=2\pi|\langle G|F_Q^A|\lambda\rangle|^2$.  To calculate
$S_{AA}(Q,\omega)$ for $H_{\theta}$ we can employ special methods at the
parameter values $\theta=\pm \pi/4$ where it is completely
integrable\cite{Takh82} or general methods otherwise.  One suitable general
method in the present context is the recursion method in combination with a
finite-size continued-fraction analysis.\cite{VM94}

Here we introduce four different fluctuation operators $F_{Q}^{A}$. The local
operators $A_n$ from which they are constructed are listed in Table~\ref{tab:2}.

\noindent
$\bullet~~$
The {\it spin} fluctuations, probed by $F_Q^S$, represent N{\'e}el order
  parameter fluctuations for $Q\!=\!\pi$. They are expected to be strongest at the
  critical point $\theta\!=\!-\pi/4$, where the $Q\!=\!\pi$ excitations are
  gapless.\\
$\bullet~~$
The {\it dimer} fluctuations, probed by $F_Q^D$, are also expected to be
  strongest (for $Q\!=\!\pi$) at $\theta\!=\!-\pi/4$, which marks the onset of dimer
  LRO.\\
$\bullet~~$
The {\it trimer} fluctuations, probed by $F_Q^T$, are constructed from
  projection operators $P_n^T$ onto local trimer states $|[1,2,3]\rangle$. The state
  $|[1,2,3]\rangle$ happens to be the ground state of $H_{\theta}$ with $N=3$ for
  $\arctan\frac{1}{3}\leq\theta\leq\pi/2$, which was interpreted as suggesting that a
  trimerized phase might exist for $N\to\infty$.\\
$\bullet~~$
The {\it center} fluctuations, probed by $F_Q^Z$, are constructed from a
modified spin operator and tune into existing period-three $(+,0,-)$ or
$(-,0,+)$ patterns of local spin states. Finite-$N$ data indicate that for
$\pi/4\leq\theta\leq\pi/2$, the fluctuations probed by $F_{2\pi/3}^Z$ are the
strongest of all the ones listed.

\vspace*{-3mm}
\begin{table}[htb]
 \caption{Local operators $A_n$ which define the four
   fluctuation operators $F_{Q}^{A}$ via \eqref{eq:FqA} and the four order
   parameters $P_{A}$ via~\eqref{eq:opa}.}
\label{tab:2} 
  \begin{tabular}{l} \hline \\
spin:\quad
{$\displaystyle S_n \equiv \left(
                \begin{array}{ccc}
                        1 & 0 & 0 \\
                        0 & 0 & 0 \\
                        0 & 0 &-1
                \end{array}     
                        \right)_n$}, \qquad
center:\quad
{$\displaystyle Z_n\equiv
        \left( \begin{array}{ccc} 
                        e^{i2\pi/3} & 0 & 0 \\ 
                        0 & 1 & 0 \\         
                        0 & 0 & e^{-i2\pi/3} 
                        \end{array} \right)_n$}
\vspace*{0.2cm} \\ \hline \\
dimer:\quad
{$\displaystyle D_n \equiv {\bf S}_n \cdot {\bf S}_{n+1} -
        \langle {\bf S}_n \cdot {\bf S}_{n+1} \rangle$}
\vspace*{0.2cm} \\ \hline \\
trimer:\quad
{$\displaystyle T_n \equiv P_n^T - \langle P_n^T \rangle, \quad
P_n^T \equiv |[n,n+1,n+2]\rangle \langle [n,n+1,n+2] |$}, \\ 
{$\displaystyle |[1,2,3]\rangle \equiv \frac{1}{\sqrt{6}}  
        \left(|+0-\rangle + |0-+\rangle + |-+0\rangle  -|-0+\rangle - |0+-\rangle - |+-0\rangle 
    \right)$}
  \\ \hline
  \end{tabular}
\end{table}

The order parameters associated with the four fluctuation operators defined
by Eq.~(\ref{eq:FqA}) and the entries of Table~\ref{tab:2} can be written in the form
\begin{equation}\label{eq:opa} 
P_A = \frac{1}{N}\sum_{n=1}^N e^{iQ_An}A_n,
\end{equation}
where $Q_S=Q_D=\pi$, and $Q_T=Q_Z=2\pi/3$. Each order parameter $P_{A}$ has a
set of eigenvectors $|\Phi_k^A\rangle, k=1,2,\ldots,K$ which represent the
associated LRO in its purest form. The degree of degeneracy is $K=2$ for the
N{\'e}el and dimer states, $K=3$ for the trimer states, and $K=6$ for center states.\cite{SMK+98}

The physical vacuum chosen here is very unlike any of the states $|\Phi_{k}^{A}
\rangle$. Within the Haldane phase, at the parameter value
$\theta_{VBS}=\arctan\frac{1}{3}$, the ground state of
$H_\theta$ is a realization of the valence-bond solid (VBS) wave
function,\cite{AKLT87} which is non-degenerate and in which the
N{\'e}el, dimer, trimer, and center ordering tendencies are all imperceptibly
weak.  In the VBS state, the spin 1 at each lattice site is expressed as a
spin-1/2 pair in a triplet state. The singlet-pair forming valence bond involves
one fictitious spin 1/2 from each of two neighboring lattice sites. The VBS
state, which has total spin $S_{T}=0$, can then be regarded as a chain of
valence bonds linking successive spin-1/2 pairs in this manner.

The topological LRO present in the Haldane phase and known to be strongest in
the VBS state provides an environment, where a specific kind of elementary
excitations can propagate freely and where the associated stationary states form
a branch with well defined dispersion. We now probe these elementary excitations
and composites thereof from different angles by the four fluctuation operators
$F_Q^A$ defined in Table~\ref{tab:2}.

Only two of the $F_Q^A$ are fully rotationally invariant,
$[F_Q^D,S_T^i]\!=\![F_Q^T,S_T^i]\!=\!0$ for $i\!=\!x,y,z$. This produces the
selection rules $\Delta S_T\!=\!0$ in the dynamic dimer and trimer structure
factors.  The corresponding selection rules for the dynamic spin and center
structure factors are $\Delta S_T\!=\!0,1$ and $\Delta S_T\!=\!0,1,2$,
respectively, with the further restriction that transitions between singlets
($S_T\!=\!0$ states) are prohibited.  At the VBS point, $S_{DD}(Q,\omega)$ and
$S_{TT}(Q,\omega)$ thus couple exclusively to the $S_T=0$ excitation spectrum,
and $S_{SS}(Q,\omega)$ exclusively to the $S_T=1$ excitation spectrum, whereas
$S_{ZZ}(Q,\omega)$ couples to the $S_T=1$ and $S_T=2$ spectra.

\begin{figure*}[htb]
\centerline{\hspace*{5mm}\epsfig{file=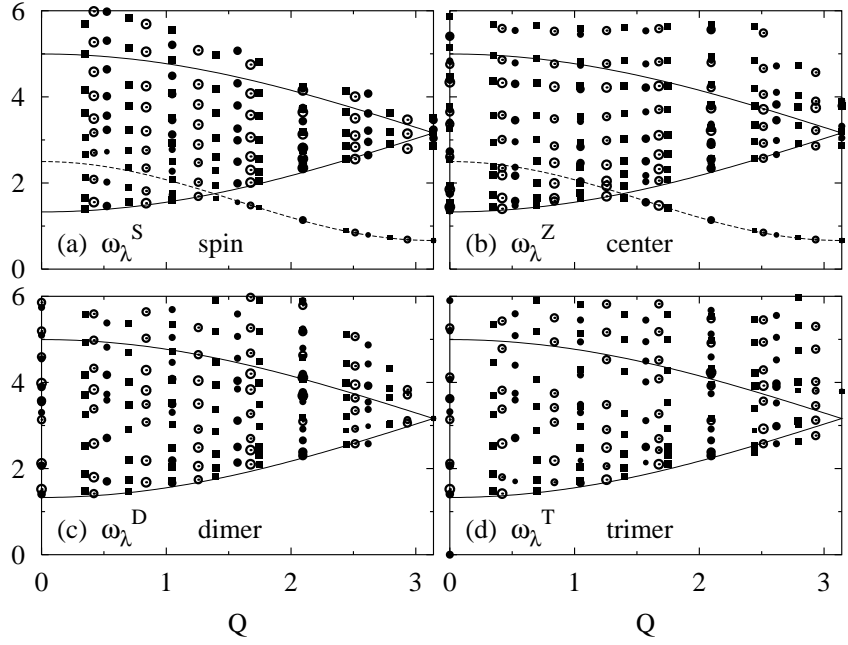,width=9.2cm,angle=-90}}
\vspace*{-8mm}
\caption{Frequency $\omega_\lambda^A$ versus wave number $Q$ of the finite-$N$
  excitations [$N\!=\!12$ ({\Large$\bullet$}), $N\!=\!15$ ({\Large$\circ$}),
  $N\!=\!18$ ($\blacksquare$)] which carry most of the spectral weight in the
  $T\!=\!0$ dynamic structure factors $S_{AA}(Q,\omega)$ for the spin, center,
  dimer, and trimer fluctuations of $H_\theta$ with $J\!=\!1$ at
  $\theta=\arctan{1\over 3}$. The three sizes of symbols used distinguish
  excitations with relative spectral weight $w_\lambda^A\equiv
  W_{\lambda}^{A}/\langle F_{Q}^{A}{F_{Q}^{A}}^{\dagger}\rangle$ in the ranges
  $w_\lambda^A \geq 0.5$ (large), $0.5 \!>\! w_\lambda^A \!\geq\! 0.1$ (medium),
  $0.1 > w_\lambda^A \geq 0.001$ (small). The dashed lines represent 1-magnon
  dispersions and the solid lines 2-magnon continuum boundaries as explained in
  the text.}\label{fig:1} 
\vspace*{-7mm}
\end{figure*}

In Fig.~\ref{fig:1} we display $\omega_\lambda^A$ versus $Q$ of the dynamically relevant
spin, center, dimer, and trimer excitation spectra as obtained from the
finite-size continued-fraction analysis.\cite{SMK+98} 
The filtered access to the spectrum afforded by the four fluctuation operators
gives us valuable clues about the nature of the elementary excitations that
thrive in the VBS environment. The low-frequency region at $Q\gtrsim \pi/2$ in the spin
and center spectra is dominated by a branch of states with $S_T=1$.  They carry
more than 95\% of the spectral weight in $S_{SS}(Q,\omega)$ and $S_{ZZ}(Q,\omega)$.
These triplets, which have been named {\it magnons}, remain invisible in the
dimer and trimer spectra. Only composites of the magnon states may be observable
in $S_{DD}(Q,\omega)$ and $S_{TT}(Q,\omega)$.  

A dispersion of the general form $ \omega_M(Q) \!=\! J(a + b\cos Q) $ for the
1-magnon branch can be inferred from the single-mode approximation of
$S_{SS}(Q,\omega)$.\cite{SMK+98,AKLT87} Under the assumption that in the VBS
environment the magnons are weakly interacting point particles, we can expect
the existence of three kinds of 2-magnon scattering states formed by pairs of
1-magnon triplets: states with $S_T\!=\!1$, which contribute to
$S_{SS}(Q,\omega)$ and $S_{ZZ}(Q,\omega)$, states with $S_T\!=\!0$, which
contribute to $S_{DD}(Q,\omega)$ and $S_{TT}(Q,\omega)$, and states with
$S_T\!=\!2$, which are observable in $S_{ZZ}(Q,\omega)$ only. Free 2-magnon
states form a two-parameter continuum $\omega_{2M}(k,Q) \equiv \omega_M(Q/2-k) +
\omega_M(Q/2+k)$ in $(Q,\omega)$-space. The resulting continuum boundaries are $
\omega_\pm(Q)=2J[a\pm b\cos(Q/2)] $.

The predicted coalescence of the 2-magnon continuum into one spectral line at
$Q=\pi$ follows from the symmetry property $\omega_M(Q) + \omega_M(\pi-Q) =
\mathit{const}$ of the magnon dispersion.  Interestingly, the finite-$N$ dimer
spectrum does indeed collapse into a single spectral line at the $N$-independent
excitation energy $\omega_D=\sqrt{10}J$, which carries all the spectral weight
in $S_{DD}(Q,\omega)$.

We now use the exact 2-magnon excitation energy, $\omega_\pm(\pi)=\omega_D$, and
the extrapolated value, $\omega_M(\pi)=0.66433(2)J$, of the 1-magnon excitation
gap to fit the parameters $a,b$. The resulting values, $a\simeq1.581,
b\simeq0.917$, used in $\omega_{M}(Q)$ and $\omega_{\pm}(Q)$ yield the dashed
and solid lines in Fig.~\ref{fig:1}.\cite{SMK+98}

Both the 1-magnon dispersion and the 2-magnon spectral threshold are in very
good agreement with all finite-$N$ data shown. Hence the magnon interaction is
very weak at the bottom of the 2-magnon region in all three $S_T$ subspaces.
The finite-$N$ data spilling out on top of the shaded areas in Fig.~\ref{fig:1}
suggest that at higher energies, the magnon interaction is repulsive, more
strongly so in the $S_T=0,2$ subspaces than in the $S_T=1$ subspace.  

This raises the possibility that bound 2-magnon states split off the top of the
2-magnon continuum of 2-magnon scattering states in the $S_T=0,2$ subspaces.
The comparison of panels (a) and (b) at frequencies $3\lesssim\omega/J\lesssim
5$ indeed suggests that the dynamically relevant finite-$N$ excitations are
arranged in contrasting patterns. In panel (a) we have an arrangement of points
which is typical of a two-parameter continuum. As $N$ increases, more points are
added and spread roughly evenly along the frequency axis. In panel (b), by
contrast, the data points are arranged in branches with an almost
$N$-independent separation, which is characteristic for branches of bound
states. The spin-1 compounds $Ni(C_{2}H_{8}N_{2})_{2}NO_{2}ClO_{4}$ (NENP) and
$Ni(C_{3}H_{10}N_{2})_{2}N_{3}(ClO_{4})$ (NINAZ), while not physical realizations
of $H_{\theta}$ directly, nevertheless realize situations where the spectrum of
collective excitations as probed by inelastic neutron scattering\cite{MBR+92}
shares major features with the magnons excited from the VBS state.
%
\section{Magnons Excited from  Ferromagnetic State}\label{sec:3}
%
We now turn to a more quantitative discussion of the difference between
scattering states and bound states in the context of completely integrable
situations, namely for the 1D $s=\frac{1}{2}$ Heisenberg ferromagnet:
\begin{equation}\label{eq:HF}
  H_F = -J\sum_{n=1}^N {\bf S}_n \cdot {\bf S}_{n+1}.
\end{equation}
The ground state is $(N\!+\!1)$-fold degenerate. We select one of the ground-state
eigenvectors, $|F\rangle \equiv |\!\uparrow\uparrow\cdots\uparrow\rangle$, as the
physical vacuum for an exploration of collective excitations.  As in the VBS
case discussed in Sec.~\ref{sec:2}, the spin fluctuation operator probes
transitions between the vacuum state and a branch of 1-magnon states. However,
here the magnon dispersion is gapless: $E\!-\!E_F\!=\!J(1\!-\!\cos k)$.

Unlike in the VBS case, the 1-magnon states of $H_{F}$ are located in a separate
invariant subspace. Only the 1-magnon states can contribute to the spin
fluctuations.  In the expression
\begin{equation}
  \label{eq:3}
  |k\rangle\equiv S_{k}^{-}|F\rangle, \quad  
  S_{k}^{-} \equiv \frac{1}{\sqrt{N}} \sum_{n=1}^{N}e^{ikn}S_{n}^{-}
\end{equation}
for the 1-magnon eigenvectors, the spin fluctuation operator plays the role of a
magnon creation operator. That is not to say $S_{k}^{-} $ is a true magnon
creation operator. Multi-magnon superpositions, i.e. the states
$S_{k_{1}}^{-} \cdots S_{k_{r}}^{-} |F\rangle$,  are a redundant set of
non-orthogonal and non-stationary states, which are subject to two kinds of
interactions:\cite{Dyso56}

\noindent
$\bullet~~$
The {\it kinematical} interaction is caused by the restriction on the
  number of reversed spins at one lattice site.\\
$\bullet~~$
The {\it dynamical} interaction is caused by the off-diagonal part of $H_F$
  in the basis of multi-magnon superpositions.
  
This distinction is quite natural in the framework of the Bethe
ansatz\cite{Beth31} as we shall see. The Bethe ansatz is an exact method for
the calculation of eigenvectors and eigenvalues of completely integrable
quantum many-body systems.  The Bethe wave function of any eigenstate in the
subspace with $r\equiv N/2-S_T^z$ reversed spins relative to the magnon
vacuum,
\begin{equation}\label{eq:psir}
 |\psi\rangle = \sum_{1\leq n_1<\ldots<n_r\leq N} a(n_1,\ldots,n_r)
 S_{n_1}^-\cdots S_{n_r}^-|F\rangle.
\end{equation}
has coefficients of the form
\begin{eqnarray}\label{eq:bar}
 a(n_1,\ldots,n_r) =
 \sum_{{\cal P}\in S_r}
 \exp && \left( i\sum_{j=1}^r k_{{\cal P} j}n_j
 + \frac{i}{2}\sum_{i<j}^{r} \theta_{{\cal P}i{\cal P}j} \right)
\end{eqnarray}
determined by $r$ magnon momenta $k_i$ and one phase angle
$\theta_{ij}=-\theta_{ji}$ for each magnon pair. The sum ${\cal P}\in S_r$ is
over the permutations of the labels $\{1,2,\ldots,r\}$.  The consistency
requirements for the coefficients $a(n_1,\ldots,n_r)$ inferred from the
eigenvalue equation $H|\psi\rangle=E|\psi\rangle$ and the requirements imposed
on the same coefficients by translational invariance can be cast in a set of
equations for the momenta $k_i$ and phase angles $\theta_{ij}$:
\label{eq:ba}
\begin{equation}\label{eq:ba1}
 2\cot\frac{\theta_{ij}}{2} = \cot\frac{k_i}{2} - \cot\frac{k_j}{2},\quad
 Nk_i = 2\pi\lambda_i + \sum_{j\neq i}^{r} \theta_{ij}.
\end{equation}
Every solution of these equations is specified by a set of $r$ Bethe quantum
numbers $\lambda_i \in \{1,2,\ldots,N-1\}$. Given a solution, the energy and wave number
of the state it describes are
\begin{equation}\label{eq:Ekr}
 E-E_F = J\sum_{j=1}^r(1-\cos k_j), \quad  k=\frac{2\pi}{N}\sum_{i=1}^r \lambda_i.
\end{equation}
In the subspace with $r\!=\!1$, we thus recover all $N$ 1-magnon states, one for
each of the allowed values of $\lambda_1$. There exist $N(N\!+\!1)/2$ distinct
2-magnon superpositions of the 1-magnon states thus identified. However, this
set of states must be accommodated in the $r=2$ subspace, whose dimensionality is
only $N(N\!-\!1)/2$. Inspection shows that of the $N(N\!+\!1)/2$ pairs of Bethe
quantum numbers in the allowed range $0\leq \lambda_{1} \leq \lambda_{2} \leq
N-1$, $N$ pairs do indeed not produce a solution of \eqref{eq:ba1}. The missing
solutions are a consequence of the kinematical interaction between magnons.

\begin{figure}[h]
  \begin{minipage}[h]{90mm}
 \centerline{\hspace*{8mm}\epsfig{file=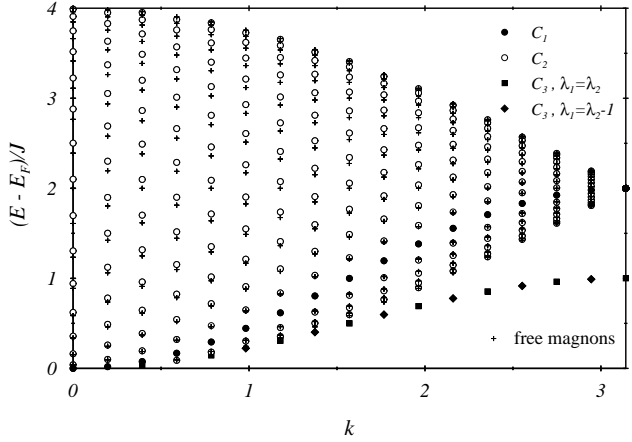,width=7cm,angle=-90}}
  \end{minipage}
  \begin{minipage}[h]{25mm}\vspace*{-8mm}
 \caption[]{Excitation energy $(E\!-\!E_F)/J$ versus wave number $k$
   of all eigenstates in the invariant subspace with $r\!=\!2$ overturned spins
   for a system with $N\!=\!32$. Also shown are the corresponding data for free
   2-magnon superpositions.}\label{fig:3}
  \end{minipage}
 \end{figure}

The consequences of the dynamical 2-magnon interaction are illustrated in
Fig.~\ref{fig:3}, which shows the complete $r\!=\!2$ spectrum $(E\!-\!E_F)/J$
versus $k$ for $k\geq0$ and $N\!=\!32$.\cite{KM97} Also shown in Fig~\ref{fig:3}
are the (fictitious) 2-magnon superpositions, where the $k_{1},k_{2}$ in
Eq.~\eqref{eq:Ekr} are replaced by all combination of 1-magnon wave numbers.
There are three classes of states.
\vspace*{-6mm}
\vspace*{-8mm}

The class $C_1$ contains $N$ states for which one of the two Bethe quantum
numbers is zero. This means that one of the magnons has zero wave number. Its
effect is a slight rotation of the magnon vacuum, in which the other magnon is
as free to propagate as in the original vacuum. There is no dynamical
interaction. All states in this class are, effectively, 1-magnon states.
  
The class $C_2$ of states has nonzero Bethe quantum numbers
$\lambda_2\!-\!\lambda_1\!\geq\! 2$.  There are $N(N\!-\!5)/2\!+\!3$ such pairs.
All of them yield a solution with real $k_1,k_2$, which makes them 2-magnon
scattering states. A measure of the dynamical magnon interaction in
Fig.~\ref{fig:3} is the vertical displacement of any true scattering state
$(\circ)$ from the nearest free-magnon pair (+).  As $N$ increases, the energy
correction diminishes for all class $C_2$ states and vanishes in the limit
$N\!\to\!\infty$. The 2-magnon scattering states and the free 2-magnon states
then form two-parameter continua with identical boundaries $E\!-\!E_F \!=\! 2J
[1\pm \cos (k/2)]$.

The class $C_3$ of states has nonzero Bethe quantum numbers which either are
equal $\lambda_2\!=\!\lambda_1$, or differ by unity
$\lambda_2\!=\!\lambda_1\!+\!1$.  There exist $2N\!-\!3$ such pairs, but only
$N\!-\!3$ pairs yield solutions of (\ref{eq:ba1}). For the class $C_3$ states,
the effects of the dynamical magnon interaction are much more prominent, and the
interaction energy does not disappear when $N\to\infty$. In Fig.~\ref{fig:3},
these states form a  branch of 2-magnon bound states with dispersion
$E\!-\!E_F \!=\! \frac{1}{2}J(1\!-\!\cos k)$ below the continuum of 2-magnon
scattering states.
\begin{figure}[h]
  \begin{minipage}{82mm}
  \centerline{\epsfig{file=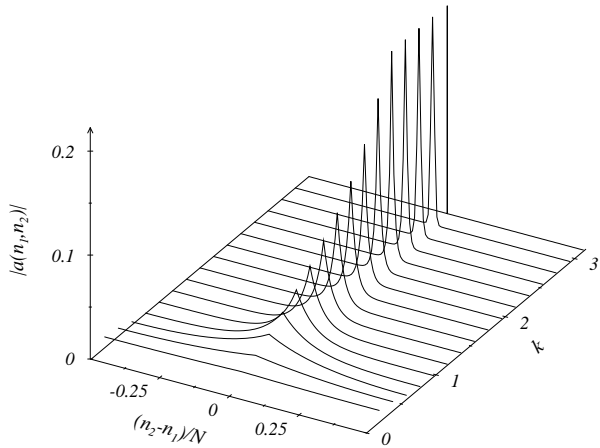,width=6.6cm,angle=-90}}
  \end{minipage}
  \begin{minipage}{25mm}
  \caption[]{Weight distribution $|a(n_1,n_2)|$ versus distance
  $n_2\!-\!n_1$ of the two down spins of class $C_3$ states at
  $k\!=\!(2\pi/N)n,\;n\!=\!4,8,\ldots,N/2$ for $N\!=\!128$.}\label{fig:4}
  \end{minipage}
\end{figure} 

The bound state character of the class $C_3$ states manifests itself in the
enhanced probability that the two flipped spins are on neighboring sites of the
lattice. This property of the wave function is best captured in the weight
distribution $|a(n_1,n_2)|$ of basis vectors with flipped spins at sites $n_1$
and $n_2$. In Fig.~\ref{fig:4} we have plotted $|a(n_1,n_2)|$ versus $n_2-n_1$
for a sequence of class $C_3$ states between $k=0$ and $k=\pi$.\cite{KM97} The
distribution is peaked at $n_2-n_1=1$. Its width is controlled by the imaginary
parts of $k_1,k_2$ in (\ref{eq:bar}).  The smallest width is observed in the
bound state at $k=\pi$. In this case, all coefficients with $n_2\neq n_1+1$ are
zero, which implies that the two down spins are tightly bound together and have
the largest binding energy.

The width of the distribution $|a(n_1,n_2)|$ increases as $k$ decreases, and the
binding of the two down spins loosens. For finite $N$, the Bethe ansatz
solutions switch from complex to real when the distribution has acquired a
certain width.  In scattering states the distribution $|a(n_1,n_2)|$ is always
broad and tends to oscillate wildly.  Some scattering states have a smooth
distribution with a maximum for $n_2=n_1+N/2$, when the two down spins are
farthest apart.  The formation of bound states and scattering states of
elementary excitations exist in many different contexts. But only in rare cases
such as this one can they be investigated on the level of detail presented here.
%
\section{Spinons Excited from Spin-Fluid State}\label{sec:4}
%
Turning our attention to the Heisenberg antiferromagnet, we consider the
Hamiltonian $H_{A}\!\equiv\! -H_{F}$ with $H_{F}$ defined in Eq.~\eqref{eq:HF}.  All
the eigenvectors remain the same, but the energy eigenvalues have the opposite
sign. The magnon vacuum $|F\rangle$ now is at the top of the excitation
spectrum.  The ground state $|A\rangle$ of $H_{A}$ is located in the invariant
subspace with $S_T^z=0$. This subspace also contains the two N{\'e}el states
$|{\mathcal N}_1\rangle \equiv
|\!\uparrow\downarrow\uparrow\cdots\downarrow\rangle, |{\mathcal N}_2\rangle
\equiv |\!\downarrow\uparrow\downarrow\cdots\uparrow\rangle$, which are not
eigenstates of $H_{A}$.  In the framework of the Bethe ansatz, $|A\rangle$ can
be obtained from $|F\rangle$ by exciting $r=N/2$ magnons with momenta $k_i$ and
(negative) energies $-J(1-\cos k_i)$. The Bethe quantum numbers for this state
are $\{\lambda_i\}_A = \{1, 3, 5, \ldots, N-1\}$.

For reasons of computational convenience we rewrite the Bethe ansatz equations
(\ref{eq:ba1}) in terms of the variables $z_i\equiv\cot(k_i/2)$:
\begin{equation}\label{eq:baez}
N\arctan z_i = \pi I_i + \sum_{j\neq i}\arctan\left(\frac{z_i-z_j}{2}\right),
\quad i=1,\ldots,r.
\end{equation}
The associated Bethe quantum numbers $-N/2<I_i\leq N/2$ are integers for odd $r$
and half integers for even $r$. The relation between the sets $\{\lambda_i\}$
and $\{I_i\}$ depends on the configuration of the solution $\{z_i\}$ in the
complex plane.  Given the solution $\{z_1,\ldots,z_r\}$ of Eqs.~(\ref{eq:baez})
for a state specified by $\{I_1,\ldots,I_r\}$, its energy and wave number are
\begin{equation}\label{eq:ekz}
\frac{E-E_F}{J} = -\sum_{i=1}^{r}\frac{2}{1+z_i^2},\quad
k = \pi r - \frac{2\pi}{N}\sum_{i=1}^r I_i,
\end{equation}
with $E_F=JN/4$. For states with real $\{z_i\}$, Eqs.~(\ref{eq:baez}) can be
converted into a convergent iterative process:
\begin{equation}\label{eq:bae-it-n}
  z_i^{(n+1)} = \tan\biggl( \frac{\pi}{N} I_i +
  \frac{1}{ N} \sum_{j\neq i}^{r}
   \arctan\bigl[\frac{z_i^{(n)}-z_j^{(n)}}{2}\bigr] \biggr)
\end{equation}
with starting values $z_i^{(1)}=\pi I_{i}/N$.  For the ground state $|A\rangle$
we have
\begin{equation}\label{eq:bqnIA}
\{I_i\}_A = \biggl\{-\frac{N}{4}+\frac{1}{2},~-\frac{N}{4}+\frac{3}{2},\ldots,
 \frac{N}{4}-\frac{1}{2}\biggr\}.
\end{equation}
High-precision solutions $\{z_i\}$ can be obtained with little computational effort. The
ground-state energy per site for $N\!=\!4096$, for example, reproduces the exact
result $(E_A\!-\!E_F)/JN \!=\!-\ln 2$ of the infinite chain to within 1 part in
a million.\cite{KHM98} The distribution of magnon momenta in the ground state
$|A\rangle$ is broad and peaked at $k_{i}=\pi$:
\begin{equation}\label{eq:rhok}
\rho_0(k_{i}) = \left[8\sin^2\frac{k_{i}}{2}\cosh
\left(\frac{\pi}{2}\cot\frac{k_{i}}{2}\right)\right]^{-1}.
\end{equation}
This state, which has obviously a very complicated structure when described in
terms of magnons, will now be configured as the physical vacuum of $H_A$ for a
different kind of elementary particle called the {\it spinon}.\cite{FT81}

\begin{figure}[htb]
  \begin{minipage}[h]{74mm}
 \centerline{\epsfig{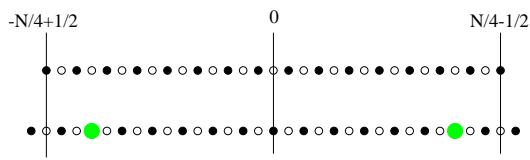}} \vspace*{-4mm}
  \end{minipage}
  \begin{minipage}[h]{42mm}
  \caption{Configurations of Bethe quantum numbers $I_i$ for the
    $N\!=\!32$ ground state and for one representative of the 2-spinon (triplet)
    excitations with $S_{T}^{z}\!=\!S_{T}\!=\!1$. Each gap in one
    $I_i$-configuration (big circles) represents a spinon.}
  \label{fig:Ii}
  \end{minipage}
\end{figure}

A useful way to characterize the new physical vacuum is through the perfectly
regular array (\ref{eq:bqnIA}) of Bethe quantum numbers as illustrated in the
first row of Fig.~\ref{fig:Ii}. 
\vspace*{-5mm}
\vspace*{-4mm}
The spectrum of $H_A$ can then be generated systematically in terms of the
fundamental excitations as characterized by elementary modifications of this
vacuum array.  

In the subspace with $S_T^z\!=\!1$, a two-parameter set of states is obtained by
removing one magnon from the state $|A\rangle$. In doing so we eliminate one of
the $N/2$ Bethe quantum numbers from the set in the first row of
Fig.~\ref{fig:Ii} and rearrange the remaining $I_i$ in all configurations over
the expanded range $|I_i| \leq \frac{1}{4}N$.  Changing $S_T^z$ by one means
that the $I_i$ switch from half-integers to integers or vice versa.  The number
of distinct configurations with $I_{i+1} \!-\! I_i \geq 1$ is $N(N\!+\!2)/8$. A
generic configuration consists of three clusters with two gaps between them as
shown in the second row of Fig.~\ref{fig:Ii}. The position of the gaps between
the $I_i$-clusters determine the momenta $\bar k_1, \bar k_2$ of the two
spinons, which, in turn, add up to the wave number of the two-spinon state
relative to the wave number of the vacuum: $Q\equiv k\!-\!k_A \!=\!  \bar
k_1\!+\! \bar k_2$.

\begin{figure}[htb]
  \begin{minipage}[h]{82mm}\vspace*{-6mm}
\centerline{~~~\epsfig{file=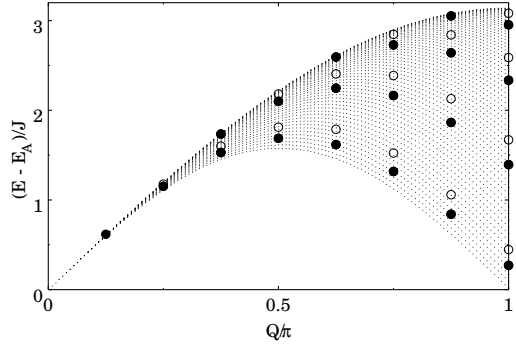,width=7.0cm,angle=0}}
  \end{minipage}
  \begin{minipage}[h]{28mm}\vspace*{8mm}
\caption{2-spinon (triplet) excitations with $S_T^z=S_T=1$ for
  $N=16$ ($\bullet$) and $N=256$ (dots). Also shown are the 2-spinon
  (singlet) excitations with $S_{T}^{z}=S_{T}=0$ for $N=16$ ($\circ$).}
\label{fig:2sp}
  \end{minipage}
\end{figure}

A plot of the 2-spinon energies $E-E_A$ versus wave number $k-k_A$ for $N=16$
as inferred via (\ref{eq:ekz}) from the solutions $\{z_i\}$ is shown in
Fig.~\ref{fig:2sp} ($\bullet$).  The dots, which represent the corresponding data
for $N=256$, produce a sort of density plot for the 2-spinon continuum which
emerges in the limit $N\to\infty$.  The exact lower and upper boundaries of the
2-spinon continuum are\cite{dCP62}
\begin{equation}\label{eq:2spb}
\omega_L(Q) = \frac{\pi}{2}J|\sin Q|,\quad
\omega_U(Q) = \pi J|\sin\frac{Q}{2}|.
\end{equation}
Like magnons, spinons carry a spin in addition to energy and momentum. Unlike
magnons, which have spin 1, spinons are spin-$1/2$ particles.  For even $N$,
where all eigenstates have integer-valued $S_T^z$, spinons occur only in pairs.
The spins $s_1,s_{2}=\pm1/2$ of the two spinons in a 2-spinon eigenstate of
$H_A$ can be combined in four different ways to form a triplet state or a
singlet state.

We have already analyzed the spectrum of the 2-spinon triplet states with
$s_1=s_2=+1/2$ $(S_T=1,\; S_T^z=+1)$.  The 2-spinon triplets with $s_1 =
s_2 =-1/2$ $(S_T=1,\; S_T^z=-1)$ are obtained from these states by a spin-flip
transformation.  One set of 2-spinon states with $s_{1}=-s_{2}$ are the triplets
with $S_{T}^{z}\!=\!0$. They are obtained via Bethe ansatz by a simple
modification of the $\{I_{i}\}$ configurations from the kind shown in the second
row of Fig.~\ref{fig:Ii} and yields an equal number of (real) solutions.  Symmetry
requires that all three triplet components $(S_T^z=0,\pm1)$ are degenerate.

The 2-spinon singlet states $(S_T^z\!=\!S_T\!=\!0)$ are characterized by one pair of
complex conjugate solutions $z_1\!=\!z_2^{\star}$ in addition to the real solutions
$z_3,\ldots,z_{N/2}$. Finding the $I_i$-configurations for a particular set of
eigenstates with complex solutions and then solving the associated Bethe ansatz
equations is, in general, delicate task.\cite{KHM98,Woyn82a} The 2-spinon
singlets for $N\!=\!16$ and $0 \!<\! Q \!\leq\! \pi$ are shown as open circles in
Fig.~\ref{fig:2sp}.  As $N$ grows large, the effect of the complex solutions
$z_1\!=\!z_2^{\ast}$ on the energy relative to that of the real solutions
$z_3,\ldots,z_{N/2}$ diminishes.  In the limit $N\to\infty$, the 2-spinon
singlets also form a continuum with boundaries (\ref{eq:2spb}). Yet the effect
of the complex solutions will remain strong for other quantities including
selection rules and transition rates.

The 2-spinon triplets play an important role in the low-temperature spin
dynamics of quasi-1D antiferromagnetic compounds such as $KCuF_{3}$,
$Cu(C_{6}D_{5}COO)_{2}\cdot 3D_{2}O$, $Cs_{2}CuCl_{4}$, and
$Cu(C_{4}H_{4}N_{2}(NO_{3})_{2})$. They are the elementary excitations of
$H_{A}$ which can be directly probed via inelastic neutron scattering.\cite{NTC+91}
The 2-spinon singlets, in contrast, cannot be excited directly from $|A\rangle$
by neutrons because of selection rules. The singlet excitations are important
nevertheless, but in a different context.  Some quasi-1D antiferromagnetic
compounds like $CuGeO_{3}$ are susceptible to a spin-Peierls transition, which
involves a lattice distortion accompanied by an exchange dimerization.\cite{AFM+96}
The operator which probes the dimer fluctuations in the ground state of $H_{A}$
couples primarily to the 2-spinon singlets and not at all to the 2-spinon
triplets.
%
\section{Spinons Excited from Antiferromagnetic State}\label{sec:5}
%
Although the computational application of the Bethe ansatz yields the exact wave
functions of the 2-spinon states for very large systems with little effort, the
very structure of the Bethe wave function makes it hard to use this knowledge
for the calculation of the transition rates pertaining to any fluctuation
operator of interest. Nevertheless, there exist ways to extract useful lineshape
information from the Bethe wave function for specific situations.\cite{KHM99}
Here we discuss an alternative approach, which exploits the higher symmetry of
$H_{A}$ in the limit $N\to\infty$, described by the quantum group $U_q(sl_2)$.
For example, the asymptotic degeneracy of the 2-spinon triplets and singlets
noted previously is attributable to this higher symmetry.

The algebraic analysis of a completely integrable spin chain employed for the
purpose of calculating dynamic structure factors for specific fluctuation
operators requires the execution of the following program:\cite{JM95}

\noindent
$\bullet~~$ 
Span the infinite-dimensional physically relevant Hilbert space in the form of a
separable Fock space of multiple spinon excitations.\\
$\bullet~~$
Generate the eigenvectors of the Hamiltonian in this Fock space by products of
spinon creation operators (so-called vertex operators) from the ground state
(physical vacuum).\\
$\bullet~~$
Determine the spectral properties (energy, momentum) of the multiple-spinon
states accessible from the ground state by the selected fluctuation operator.\\
$\bullet~~$
Express the fluctuation operator of interest in terms of vertex operators.\\
$\bullet~~$
Evaluate matrix elements of products of vertex operators as are needed for the
selected fluctuation operator in the spinon eigenbasis.\\

In the following, we outline how this program was carried out for the
calculation of the exact 2-spinon part of one dynamic structure factor for the
1D $s=\frac{1}{2}$ $XXZ$ model,
\begin{equation}\label{eq:HXXZ}
  H_{XXZ} = -J \sum_{n=-\infty}^{\infty} 
  (S_n^x S_{n+1}^x +S_n^y S_{n+1}^y + \Delta S_n^z S_{n+1}^z).
\end{equation}
At $T=0$ $H_{XXZ}$ has a ferromagnetic phase for $\Delta\geq 1$, a critical
phase (spin-fluid) for $-1 \leq \Delta < 1 $, and an antiferromagnetic phase for
$\Delta \!<\! -1$.  The algebraic analysis operates in the massive phase
stabilized by N{\'e}el LRO at $\Delta\!<\!-1$, but the isotropic limit $\Delta
\!\to\! -1^-$, which is equivalent to $H_{A}$, can be performed meaningfully.

Each spinon in the $m$-spinon eigenstate
$|\xi_m,\epsilon_m;\ldots;\xi_1,\epsilon_1\rangle_j$ is characterized by a
(complex) spectral parameter $\xi_l$ of unit length and a spin orientation
$\epsilon_l\!=\!\pm1$. The spectral properties then follow from the rules
governing the application of the translation operator and the
Hamiltonian:\cite{JM95}
\begin{eqnarray}\label{eq:txiop}
    T|\xi_m\!,\!\epsilon_m\!;\!\ldots\!;\!\xi_1\!,\!\epsilon_1\rangle_j 
&=&
    \prod_{i=1}^m\frac{1}{\tau(\xi_i)}
    |\xi_m\!,\!\epsilon_m\!;\!\ldots\!;\!\xi_1\!,\!\epsilon_1\rangle_{1\!-\!j}, \\
\label{eq:hxiop}
    H|\xi_m\!,\!\epsilon_m\!;\!\ldots\!;\!\xi_1\!,\!\epsilon_1\rangle_j &=&
    \sum_{i=1}^m e(\xi_i)
    |\xi_m\!,\!\epsilon_m\!;\!\ldots\!;\!\xi_1\!,\!\epsilon_1\rangle_j,
\end{eqnarray}
\begin{eqnarray}\label{eq:tauxi}
    \tau(\xi)
&=& 
e^{-ip(\xi)} = \xi^{-1} \frac{\theta_{q^4}(q\xi^2)}{\theta_{q^4}(q \xi^{-2})}, \quad
    e(\xi) = J\frac{1-q^2}{4q} \xi \frac{d}{d\xi} \log \tau(\xi), 
\end{eqnarray}
\begin{equation}
\label{eq:jmdef}
    \theta_x(y)  \equiv (x;x)(y;x)(x y^{-1};x),  \quad 
    (y;x) \equiv \prod_{n=0}^{\infty} (1-y x^n).
\end{equation}
Here $q$ is the deformation parameter of the quantum group $U_q(sl_2)$. Its
value is determined by the exchange anisotropy, $\Delta=(q+q^{-1})/2$.  The
twofold degenerate vacuum is represented by the vectors $|0\rangle_0,
|0\rangle_1$, which transform into each other via translation: $T|0\rangle_j =
|0\rangle_{1-j}$. In the Ising limit $\Delta\to-\infty$, they become the pure
N{\'e}el states $|\ldots\uparrow\downarrow\uparrow\downarrow\ldots\rangle,
|\ldots\downarrow\uparrow\downarrow\uparrow\ldots\rangle$.

By eliminating $\xi$ from the relations (\ref{eq:txiop}) and (\ref{eq:hxiop}) we
obtain the spinon energy-momentum relation
\begin{equation}\label{eq:e1p}
e_1(p) = I\sqrt{1-k^2\cos^2p},\quad I\equiv\frac{JK}{\pi}\sinh\frac{\pi K'}{K},
\end{equation}
which is independent of the spin orientation. The elliptic integrals $K\equiv
K(k), K'\equiv K(k')$ and their moduli $k,k'\equiv \sqrt{1-k^2}$ depend on the
anisotropy via $-q = \exp(-\pi K^\prime/K)$.
\begin{figure}[htb]
  \begin{minipage}[h]{81mm}
  \centerline{\vspace*{-9mm}~\epsfig{file=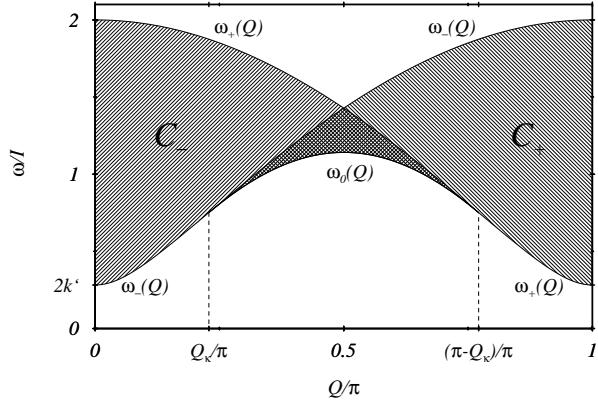,width=6.3cm,angle=-90}}
  \end{minipage}
  \begin{minipage}[h]{31mm}\vspace*{-6mm}
    \caption{2-spinon excitation spectrum for $k=0.99$
    $(\Delta\simeq -2.305)$. It consists of two partly overlapping sheets
    $\mathcal{C}_-$ and $\mathcal{C}_+$. Sheet $\mathcal{C}_+$ lies between
    $\omega_0(Q)$ and $\omega_-(Q)$ in the range $Q_\kappa \leq Q \leq \pi -
    Q_\kappa$ and between $\omega_+(Q)$ and $\omega_-(Q)$ for $\pi-Q_\kappa \leq
    Q \leq \pi$.  The interval $(0,\pi)$ represents one half of the extended
    Brillouin zone.}  \label{fig:8}
  \end{minipage}
\end{figure}


The of 2-spinon spectrum with energies $E(\xi_1,\xi_2) = e(\xi_1)+e(\xi_2)$ and
momenta $P(\xi_1,\xi_2) = p(\xi_1)+p(\xi_2)$ is then a two-parameter set with
fourfold spin degeneracy $\epsilon_1,\epsilon_2=\pm 1$.\cite{JM72} In the
isotropic limit, they are the 2-spinon triplets and singlets discussed in
Sec.~\ref{sec:4}. In a finite system, the singlet-triplet degeneracy is removed
(see Fig.~\ref{fig:2sp}). Exchange anisotropy splits up the triplet levels as
well.

In the $(Q,\omega)$ plane, each set of 2-spinon excitations forms a continuum
(see Fig.~\ref{fig:8}) of two sheets $\mathcal{C}_\pm$ with boundaries
\label{eq:MM25}
\begin{equation}
  \omega_0(Q)   = \frac{2I}{1+\kappa}\sin Q, \quad
  \label{eq:MM25a}
  \omega_{\pm}(Q) = \frac{2I}{1+\kappa}\sqrt{1+\kappa^{2}\pm 2\kappa\cos Q},
\end{equation}
where $ \kappa \equiv \cos Q_\kappa = (1-k')/(1+k') $ is a convenient anisotropy
parameter.   The evaluation of the dynamic structure factor of any
fluctuation operator $F_Q^A$, Eq.~(\ref{eq:FqA}), requires that we know exact matrix
elements $_j\langle0|A_l|\xi_m,\epsilon_m,\ldots,\xi_1,\epsilon_1\rangle_{j'}$
for the associated local operators. In principle, they can be calculated
exactly if we are able to express $A_l$ in terms of vertex operators.  In
practice, the exact results available thus far are limited to the dynamic spin
structure factors
\begin{equation}\label{eq:stfdef}
  S_{\mu\mu}(Q,\omega)=\sum_{n=-\infty}^{+\infty}
  \int_{-\infty}^{\infty} dt\; e^{i(\omega t+Qn)} 
  \langle S_n^\mu(t) S_0^\mu\rangle,\quad
  \mu = x,z.
\end{equation}
 
 In the $m$-spinon expansion $S_{zz}(Q,\omega) \!=\! S_{zz}^{(0)}(Q,\omega) \!+\!
 S_{zz}^{(2)}(Q,\omega) + \ldots$ we know the matrix elements that determine the
 leading term,\cite{JM95}
\begin{equation}\label{eq:stagm}
2{_j}\langle 0|S_n^z|0\rangle_j =
    \frac{(q^2;q^2)^2}{(-q^2;q^2)^2}(-1)^{n+j} =
    \frac{2\pi}{K}\sqrt{k'} = 2\bar{m}_z,
  \end{equation}
  yielding $ S^{(0)}_{zz}(Q,\omega) \!=\!
  4\pi^2\delta(\omega)\delta(Q-\pi)\bar{m}_z^2, $ where $\bar{m}_z$ is the
  staggered magnetization.\cite{Baxt73} No exact transition rates have been
  evaluated for the 2-spinon part $S_{zz}^{(2)}(Q,\omega)$ or any higher order
  term in the $m$-spinon expansion.

\begin{figure*}[htb]
  \centerline{\epsfig{file=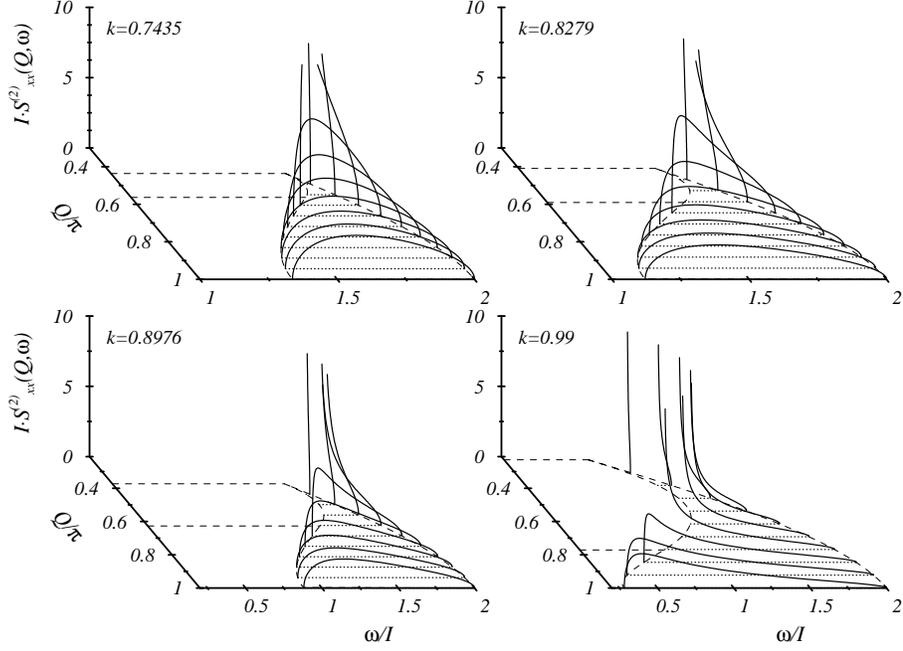,width=10.0cm,angle=-90}}
  \vspace*{-9mm}
  \caption{2-spinon dynamic structure factor $S_{xx}^{(2)}(Q,\omega)$ for
    $(Q,\omega)\in\mathcal{C_+}$ as a function of frequency for wave numbers
    $Q_\kappa \leq Q \leq \pi$ and anisotropy parameter $k=0.7435$
    ($\Delta\simeq -10$), $k=0.8279$ ($\Delta\simeq -7$), $k=0.8976$
    ($\Delta\simeq -5$), and  $k=0.99$ ($\Delta\simeq-2.305$).} 
  \label{fig:sqw4}\vspace*{-5mm}
\end{figure*}

The corresponding $m$-spinon expansion of $S_{xx}(Q,\omega)$ starts with $m\!=\!2$,
because the operator $S_n^x$ does not connect the vacuum sector with
itself. All non-vanishing matrix elements which are needed for
$S_{xx}^{(2)}(Q,\omega)$ can be expressed, as it turns out, by a single function
$X^j(\xi_2,\xi_1) \equiv {_j\langle}0|\sigma_0^+|\xi_2,-;\xi_1,-\rangle_j$, which was
determined by Jimbo and Miwa:\cite{JM95}
\begin{equation}\label{eq:Xjjm2}
    X^j(\xi_2,\xi_1)  =
 \varrho^2\frac{(q^4;q^4)^2}{(q^2;q^2)^3}
  \frac{(-q\xi_1\xi_2)^{1-j}\xi_1
  \gamma(\xi_1^2/\xi_2^2) \theta_{q^8}(-\xi_1^{-2}\xi_2^{-2}q^{4j})}%
  {\theta_{q^4}(-\xi_1^{-2}q^{3})\theta_{q^4}(-\xi_2^{-2}q^{3})},
\end{equation}
where $ (x;y;z)\equiv \prod_{n,m=0}^{\infty}\left(1-xy^{n}z^{m}\right)$ and
\label{eq:jmdef2}
\begin{eqnarray}
    \gamma(\xi)
&\equiv&
    \frac{(q^4\xi;q^4;q^4)(\xi^{-1};q^4;q^4)}%
     {(q^6\xi;q^4;q^4)(q^2\xi^{-1};q^4;q^4)}, \quad
    \varrho \equiv
  (q^2;q^2)^2 \frac{(q^4;q^4;q^4)}{(q^6;q^4;q^4)}.
\end{eqnarray} 
These ingredients yield the following expression to be evaluated:\cite{BKM98}
\begin{eqnarray}\label{eq:s2xixi}
S^{(2)}_{xx}(Q,\omega) &=& 
    \frac{1}{4} \oint\frac{d\xi_1}{2i\xi_1}\frac{d\xi_2}{2i\xi_2}
    \delta[\omega-E(\xi_1,\xi_2)] \nonumber
    \\  && \hspace{-20mm} \times 
    \{\delta[ Q +P(\xi_1,\xi_2)] |X^0(\xi_2,\xi_1)+X^1(\xi_2,\xi_1)|^2 \nonumber
    \\  && \hspace{-20mm} +  
    \delta[ Q-\pi +P(\xi_1,\xi_2)] |X^0(\xi_2,\xi_1)-X^1(\xi_2,\xi_1)|^2\}.
  \end{eqnarray}
A compact rendition of the exact result  reads\cite{BKM98}
\begin{equation}\label{eq:sxxfin}
 S_{xx}^{(2)}(Q,\omega) = \frac{\omega_0}{8I\omega}
    \left[ 1 \!+\! 
      \sqrt{\frac{\omega^{2}\!-\!\kappa^{2}\omega^{2}_0}%
        {\omega^{2}\!-\!\omega^{2}_0}}
    \right] 
    \sum_{c=\pm}  
    \frac{\vartheta_A^2(\beta_-^{c})}{\vartheta_d^2(\beta_-^c)} 
    \frac{|\tan(Q/2)|^{-c}}{W_c},
\end{equation}
where 
\begin{equation}\label{eq:Wpm}
 W_\pm= 
    \sqrt{\frac{\omega_0^4}{\omega^4} \kappa^2 
      -\left(\frac{T}{\omega^2} \pm \cos Q\right)^2},\quad
  T = \sqrt{\omega^2-\kappa^2\omega_0^2} \sqrt{\omega^2-\omega_0^2},
\end{equation}
\begin{equation}\label{eq:landmin}
   \beta_-^c(Q,\omega) = \frac{1+\kappa}{2} 
    F\left[ \arcsin\left(
        \frac{2I\omega W_c}{\kappa(1+\kappa)\omega_0^2}
      \right),\kappa \right],
\end{equation}
\begin{equation}\label{eq:thetaa}
  \vartheta_A^2(\beta) = \!\exp\left(\!-\sum_{l=1}^\infty 
      \frac{e^{\gamma l}}{l}
      \frac{\cosh(2\gamma l)\cos(t\gamma l)-1}%
      {\sinh(2l\gamma)\cosh(\gamma l)}
      \right),
\end{equation}
$\gamma = \pi K'/K$, $t\equiv 2\beta/K'$, and $\vartheta_d(x)$ is a Neville
theta function. This function is plotted in Fig.~\ref{fig:sqw4} for
$(Q,\omega)\in \mathcal{C_+}$.  $S_{xx}^{(2)}(Q,\omega)$ has a square-root
divergence at the portion $\omega_0(Q)$ of the spectral threshold. Along the
portion $\omega_+(Q)$ of the lower boundary and along the entire upper boundary
of $\mathcal{C}_+$, $S_{xx}^{(2)}(Q,\omega)$ has a square-root cusp.  The line
shapes for $(Q,\omega)\in \mathcal{C_-}$ are inferred from the symmetry property
$S_{xx}^{(2)}(\pi-Q,\omega)=S_{xx}^{(2)}(Q,\omega)$.

\begin{figure}[htb]
  \begin{minipage}[h]{82mm}
  \centerline{\vspace*{4mm}\epsfig{file=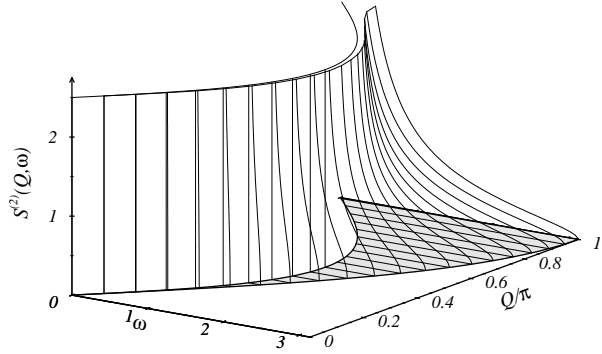,width=6.4cm,angle=-90}}
  \end{minipage}
  \begin{minipage}[h]{32mm} \vspace*{-21mm}
  \caption{Exact 2-spinon dynamic spin structure
     factor. The expression is nonzero only in the shaded region of the
     $(Q,\omega)$-plane bounded by $\omega_L(Q)$ and $\omega_U(Q)$.}
   \label{fig:isotropic}
  \end{minipage}  \vspace*{-14mm}
\end{figure}

The isotropic limit $\Delta\to-1^-$ is delicate because of its singular nature.
As the LRO in the spinon vacuum vanishes, the size of the Brillouin zone changes
from $(-\pi/2,+\pi/2)$ to $(-\pi,+\pi)$. A practical consequence of this phase
transition is that we switch our perspective from considering both sheets
$\mathcal{C}_\pm$ of 2-spinon excitations over the range $(-\pi/2,+\pi/2)$ to
considering only the sheet $\mathcal{C}_+$, now with boundaries \eqref{eq:2spb},
over the extended range $(-\pi,+\pi)$.  With these subtleties taken into
account, Eq.~(\ref{eq:sxxfin}) reduces to\cite{KMB+97}
\label{eq:sxxis}
\begin{equation}\label{eq:sxxis1}
S_{xx}^{(2)}(Q,\omega) = \frac{1}{2}[\omega_U^2(Q)-\omega^2]^{-1/2}e^{-I(t)},
\end{equation}
\begin{equation}
  \label{eq:sxxis2}
  I(t) = 
  \int_0^\infty dx \frac{\cosh 2x \cos xt -1}{x \sinh 2x \cosh x}e^x , \quad
  \frac{\pi t}{4} = 
  {\rm cosh}^{-1}\sqrt{\frac{\omega_U^2(Q)-\omega_L^2(Q)}{\omega^2-\omega_L^2(Q)}}.
\end{equation}
This function, which is plotted in Fig.~\ref{fig:isotropic}, has a square-root cusp at
$\omega_U(Q)$ and a square-root divergence with logarithmic corrections at
$\omega_L(Q)$.
\vspace*{-4mm}

%
\section{Solitons Excited from N{\'e}el State}\label{sec:solit-excit-from}
\begin{figure*}[htb]
  \begin{minipage}[h]{81mm}
  \centerline{\epsfig{file=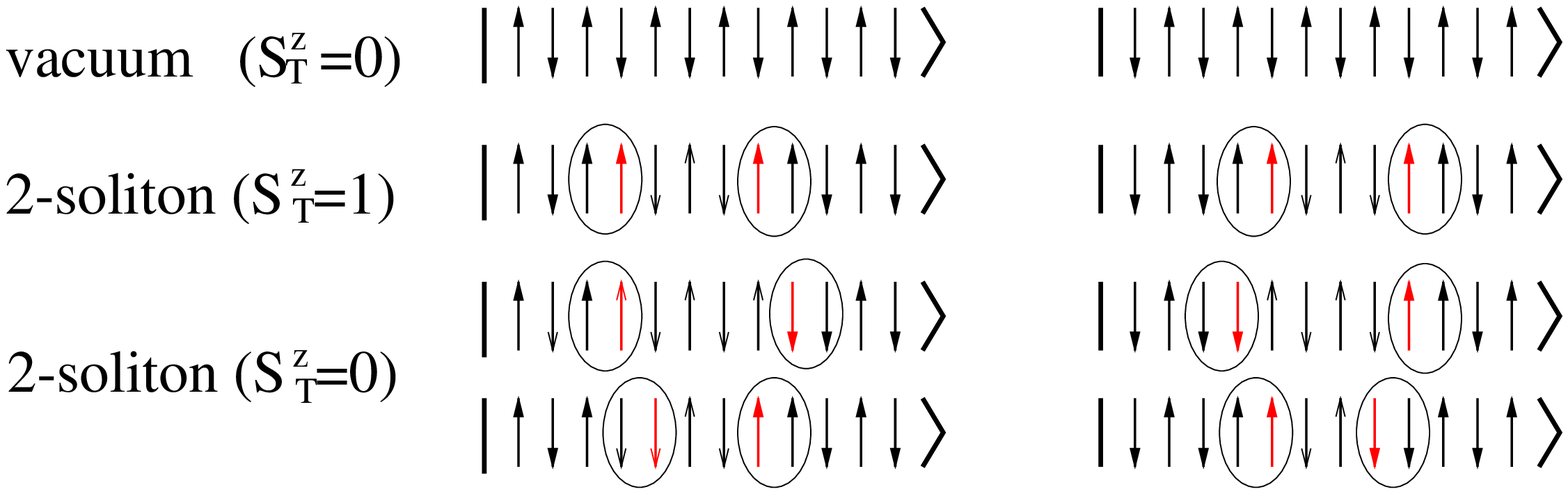,width=7.6cm}}
  \end{minipage}
  \begin{minipage}[h]{34mm}
  \caption{Local basis states from which translationally invariant N{\'e}el states
    and 2-soliton states are constructed. The latter are limiting cases of the
    2-spinon states discussed in Sec.~\ref{sec:5}.} 
  \label{fig:soliton}
  \end{minipage}
\vspace*{-5mm}
\end{figure*}

%
Near the Ising limit ($\Delta\to-\infty$), the exact result (\ref{eq:sxxfin}) can be
expanded in powers of the anisotropy parameter $\kappa$. To
leading order, we obtain
 \begin{equation}\label{eq:ising-dsf-zero}
    S_{xx}^{(2)}(Q,\omega) \to
    \frac{1}{2\cos^{2}Q}
    \sqrt{\cos^{2}Q-\left(\frac{\Omega}{\kappa}\right)^2\Theta(\kappa |\cos Q|-|\Omega|)}
 \end{equation}
 with $\Omega \equiv \omega/2I - 1$, which is identical (in that order) to the
 result obtained by Ishimura and Shiba\cite{IS80} from a first-order
 perturbation calculation about the Ising limit.  In their calculation, the two
 vacuum states are approximated by the pure N{\'e}el states (see
 Fig.~\ref{fig:soliton}).  The spinons are kink solitons (domain
 walls),\cite{Vill75} which produce two adjacent up or down spins in an
 otherwise unperturbed N{\'e}el configuration. The only states which contribute
 to $S_{xx}(Q,\omega)$ in leading order of the perturbation calculation are
 states which contain two solitons with equal spin orientation.  The states with
 solitons of opposite spin orientation dominate the spectrum of
 $S_{zz}(Q,\omega)$ except for the central peak at $\omega=0,\; Q=\pi$, which
 involves a transition between the two vacuum states.
\vspace*{-5mm}

\begin{figure}[htb]
  \begin{minipage}[h]{84mm}
  \centerline{\epsfig{file=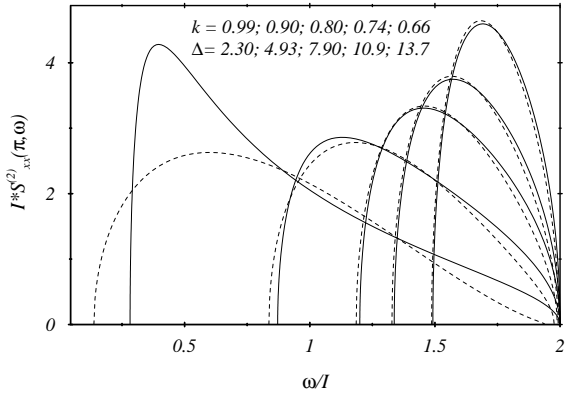,width=6.2cm,angle=-90}}
  \end{minipage}
  \begin{minipage}[h]{28mm}\vspace*{-18mm}
  \caption{Line shapes of $S_{xx}^{(2)}(Q,\omega)$ at the zone center $Q=\pi$ as 
    established by the exact result \eqref{eq:sxxfin} (solid line) and predicted
    by the first order perturbation result~\eqref{eq:ising-dsf-zero} (dashed line).  }
  \label{fig:10}
  \end{minipage}\vspace*{-10mm}
\end{figure}

The 2-soliton spectrum with boundaries $|\Omega|\leq \kappa \cos Q$, is akin to
the generic 2-spinon spectrum shown in Fig.~\ref{fig:8}, but it does not capture
any of the features that characterize the regime $Q_{\kappa}\leq Q \leq \pi
-Q_{\kappa}$ around the Brillouin zone boundary.  The limitations of the
perturbation results are much less severe near the center of the Brillouin zone.
This illustrated in Fig.~\ref{fig:10}, which compares the line shapes of Eq.
\eqref{eq:sxxfin} and \eqref{eq:ising-dsf-zero} at $Q=\pi$. Here a fairly strong
departure from the Ising limit is required before a significant difference
between the two results is produced.
\vspace*{-2mm}

Two intensively studied physical realizations of $H_{XXZ}$ at $\Delta< -1$ are
$\rm CsCoCl_3$ and $\rm CsCoBr_3$. Spectroscopic data which probe both the
quantum and thermal fluctuations of those materials are available from several
neutron scattering experiments.\cite{YHSS81}

%
\section*{Acknowledgments}
%
Financial support from the URI Research Office (for G.M.) and from the DFG
Schwerpunkt \textit{Kollektive Quantenzust{\"a}nde in elektronischen 1D
  {\"U}bergangsmetallverbindungen} (for M.K.) is gratefully acknowledged.

\end{document}